\documentclass[10pt,twocolumn]{article}
\usepackage{graphicx} 
\graphicspath{{figures}} 
\usepackage[font=small,labelfont=bf]{caption} 

\usepackage{xcolor}
\definecolor{black}{HTML}{212427}
\definecolor{blue}{HTML}{0563C1}
\definecolor{brightred}{HTML}{FF0000} 
\DeclareUnicodeCharacter{2212}{-}
\usepackage{hyperref,nameref}
\hypersetup{colorlinks=true,allcolors=blue}
\newcommand{\rref}[2]{\hyperref[#1]{\ref{#1}#2}} 

\color{black}
\usepackage[margin=0.67in]{geometry} 

\usepackage[switch]{lineno}   
\usepackage{amsmath,amssymb}  
\usepackage{etoolbox} 
\newcommand*\linenomathpatch[1]{%
  \cspreto{#1}{\linenomath}%
  \cspreto{#1*}{\linenomath}%
  \csappto{end#1}{\endlinenomath}%
  \csappto{end#1*}{\endlinenomath}%
}
\linenomathpatch{equation}
\linenomathpatch{gather}
\linenomathpatch{multline}
\linenomathpatch{align}
\linenomathpatch{alignat}
\linenomathpatch{flalign}

\usepackage{fancyhdr,lastpage} 
\usepackage{titlesec}
\titleformat{\section}{\Large\bfseries}{}{0mm}{}
\titleformat{\subsection}{\bfseries}{}{0mm}{}
\titlespacing{\section}{0pt}{\baselineskip}{0pt}
\titlespacing*{\section}{0pt}{\baselineskip}{0pt}
\titlespacing*{\subsection}{0pt}{\baselineskip}{0pt}
\usepackage{float}
\renewcommand{\t}[1]{\text{#1}} 

\usepackage[
  backend=biber,        
  autocite=superscript, 
  sorting=none,         
  style=numeric-comp,   
  maxnames=100,         
  minnames=100,         
  giveninits=false,     
  autopunct=false,      
  doi=true,             
  hyperref=true         
]{biblatex}
\DeclareFieldFormat{labelnumberwidth}{\;[#1]\!\!\!\!} 
\renewbibmacro{in:}{} 
\AtNextBibliography{\small} 
\AtEveryBibitem{\clearfield{pages}} 
\AtEveryBibitem{\clearfield{volume}} 
\AtEveryBibitem{\clearfield{number}} 
\AtEveryBibitem{\clearfield{month}} 
\AtEveryBibitem{\clearfield{eprint}} 
\AtEveryBibitem{\clearfield{issn}} 
\AtEveryBibitem{
    \clearfield{urlyear}
    \clearfield{urlmonth}
}

\usepackage[shortcuts, abbreviations,]{glossaries-extra}

\newabbreviation{SRO}{SRO}{short-range order}
\newabbreviation{SS}{SS}{solid solution}
\newabbreviation{fcc}{fcc}{face-centered cubic}
\newabbreviation{bcc}{bcc}{body-centered cubic}
\newabbreviation{hcp}{hcp}{hexagonal close-packed}
\newabbreviation{WC}{WC}{Warren-Cowley}
\newabbreviation{1CP}{1CP}{first coordination polyhedron}
\newabbreviation{KL}{KL}{Kullback-Leibler}
\newabbreviation{ML}{ML}{machine learning}
\newabbreviation{MC}{MC}{Monte Carlo}
\newabbreviation{1NN}{1NN}{first nearest neighbors}
\newabbreviation{MLP}{MLP}{machine learning potential}
\newabbreviation{DFT}{DFT}{density functional theory}

\glsdisablehyper 

\usepackage[]{cleveref} 
\creflabelformat{equation}{#2\textup{#1}#3} 

\usepackage{tabularx}
\usepackage{booktabs}
\newcolumntype{Y}{>{\raggedleft\arraybackslash}X} 
\usepackage{multirow}

\begin{document}

\twocolumn[
  \begin{center}
	\large
    \textbf{Machine learning potentials for modeling alloys across compositions}
  \end{center}
  Killian Sheriff$^1$,
  Daniel Xiao$^1$,
  Yifan Cao$^1$,
  Lewis R. Owen$^2$,
  and Rodrigo Freitas$^1${\footnotemark[1]} \\
  $^1$\textit{\small Department of Materials Science and Engineering, Massachusetts Institute of Technology, Cambridge, MA, USA} \\
  $^2$\textit{\small Department of Materials Science and Engineering, University of Sheffield, UK} \\
  
  {\small Dated: \today}
  \vspace{-0.15cm}
  \begin{center}
	\textbf{Abstract}
 \end{center}
    \vspace{-0.35cm}
Materials properties depend strongly on chemical composition, i.e., the relative amounts of each chemical element. Changes in composition lead to entirely different chemical arrangements, which vary in complexity from perfectly ordered (i.e., stoichiometric compounds) to completely disordered (i.e., solid solutions).
Accurately capturing this range of chemical arrangements remains a major challenge, limiting the predictive accuracy of machine learning potentials (MLPs) in materials modeling.
Here, we combine information theory and machine learning to optimize the sampling of chemical motifs and design MLPs that effectively capture the behavior of metallic alloys across their entire compositional and structural landscape.
The effectiveness of this approach is demonstrated by predicting the compositional dependence of various material properties --- including stacking-fault energies, short-range order, heat capacities, and phase diagrams --- for the AuPt and CuAu binary alloys, the ternary CrCoNi, and the TiTaVW high-entropy alloy. 
Extensive comparison against experimental data demonstrates the robustness of this approach in enabling materials modeling with high physical fidelity.
\vspace{0.4cm}
] 
{
  \footnotetext[1]{Corresponding author (\texttt{rodrigof@mit.edu}).}
}

\begin{figure}[!b]
  \centering
  \includegraphics{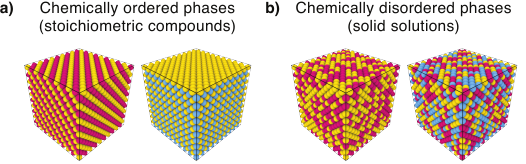}
    \caption{\label{fig:figure_1} \textbf{Chemical complexity of phases in metallic alloys.} \textbf{a)} Stoichiometric compounds exhibit ordered atomic arrangements with fixed elemental ratios. \textbf{b)} Solid solutions are chemically disordered phases with flexible composition and greater chemical complexity.}
\end{figure}

\noindent The properties of materials are heavily influenced by their chemical composition, i.e., the relative amounts of each constituent chemical element. Changes in composition lead to striking variations in the chemical arrangements of atoms that drive macroscopic behavior. 
For example, in stoichiometric compounds (fig.~\rref{fig:figure_1}{a}), the elements combine in fixed and precise ratios, and are organized in a periodic and highly ordered phase with little chemical complexity. 
In contrast, solid solutions (fig.~\rref{fig:figure_1}{b}) are phases in which chemical elements are distributed in a nearly disordered manner, allowing greater flexibility in composition. 
These differences in atomic arrangements often translate into distinct macroscopic properties. 
For example, in the CuAu alloy, the ordered intermetallic phase Cu$_3$Au is brittle, while the disordered CuAu solid solution is ductile\autocite{anderson2017theory}.

Although machine learning potentials (MLPs) have demonstrated remarkable predictive accuracy for stoichiometric compounds\autocite{neumann2024orb, yang2024mattersim, TANG2022118217,thorn_machine_2023}, extending this success to solid solutions has proven challenging\autocite{cao2024capturing, roadmap, sheriff2023quantifying, Song2024}. 
Addressing this limitation is important because solid solutions are prevalent across the materials design space. For example, they are the dominant phase over wide compositional ranges in the phase diagrams of metallic alloys\autocite{ASMhandbook} (e.g., fig.~\ref{fig:figure_5}). Notably, much of the widespread interest in high-entropy materials stems from the chemical complexity of solid solutions, and the flexibility they enable in designing mechanical and functional properties\autocite{Han2024, george_high-entropy_2019, Oses2020}.

The main challenge impeding the development of high-accuracy MLPs for solid solutions is the large variability in the chemical arrangement of atoms in this phase\autocite{sheriff2023quantifying,sheriff2024chemicalmotif}, which is immensely amplified when changes in chemical composition are taken into account. Generating an MLP training set that faithfully represents this chemical complexity is notoriously difficult\autocite{roadmap}. The current state-of-the-art strategy\autocite{cao2024capturing, musa2024accelerating} employs atomistic configurations obtained from Density-Functional-Theory-based Monte Carlo (DFT-MC) trajectories. While physically grounded, this approach is prohibitively expensive---often requiring more than 100,000 CPU hours to create a training set for a single composition\autocite{cao2024capturing}. Furthermore, this approach is limited to equilibrium states, which is a critical drawback because MLPs trained on equilibrium datasets generalize poorly to out-of-distribution nonequilibrium configurations commonly encountered during materials synthesis and processing\autocite{islam2024nonequilibriumchemicalshortrangeorder, sheriff2024simultaneous, Merchant2023, neumann2024orb}. Emerging approaches such as active learning\autocite{podryabinkin2023mlip3activelearningatomic, Tan2023,pathrudkar2024electronicstructurepredictionmedium,Vandermause2022}, though efficient for optimizing first-principles datasets, often lack objective functions capable of capturing chemical complexity, relying instead on system representations that have been shown to obscure the relevant variations in local chemical environments\autocite{sheriff2023quantifying}.

Here, we demonstrate the design of MLPs that accurately model metallic alloys across their compositional and structural landscape. 
This is achieved by combining information theory with geometric deep learning to construct training datasets optimized to capture chemical complexity through motif-based sampling. 
Extensive comparisons against experimental data show that this approach produces high-fidelity models of complex material systems --- ranging from simple binary systems to high-entropy alloys.


\section{Results}

\subsection{Performance of universal ML potentials on solid solutions}

\begin{figure}[!b]
  \centering
  \includegraphics{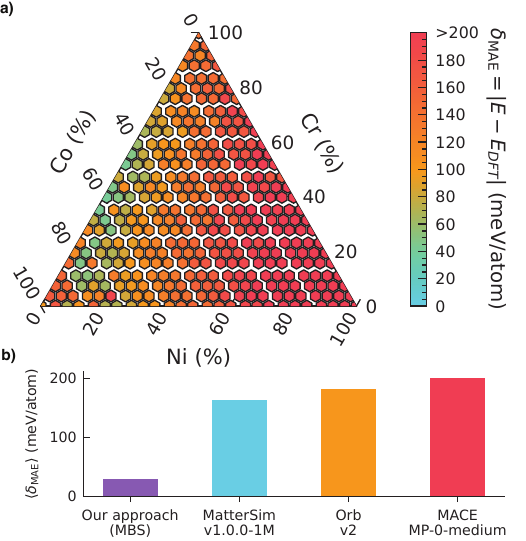}
  \caption{\label{fig:figure_2}
    \textbf{Error of universal machine learning potentials (uMLPs) on solid solutions.}
    \textbf{a}) Energy mean absolute error of MatterSim-v1.0.0-1M\autocite{yang2024mattersim} across the composition space of CrCoNi random fcc solid solution. 
    \textbf{b}) Comparison of error averaged across the compositional space for our motif-based sampling (MBS) approach and various uMLPs\autocite{yang2024mattersim,neumann2024orb,batatia2023foundation} for the same alloy.
  }
\end{figure}

Universal machine learning potentials (uMLPs) are a class of models trained on broad datasets encompassing a wide range of chemistries, compositions, and crystal structures. These models aim to achieve robust transferability across diverse material systems. However, this promise of universality often breaks down in chemically disordered phases. For example, consider fig.~\rref{fig:figure_2}{a}, which shows the accuracy of the MatterSim\autocite{yang2024mattersim} uMLP across the composition space of the CrCoNi solid solution --- a prototypical high-entropy alloy (see the Methods section ``Training and testing datasets'' for dataset details). 
The mean absolute error (MAE, $\delta_\t{MAE} =  |E-E_{\t{DFT}}|$) between predicted and reference DFT energies reaches up to 4,500\,meV/atom, with as much as 10,861\% variation across compositions (i.e., two orders of magnitude). This compositional sensitivity implies that the reliability of predicted physical properties can vary dramatically with composition.

This behavior is not unique to MatterSim. Other state-of-the-art uMLPs --- namely, Orb\autocite{neumann2024orb} and MACE\autocite{batatia2023foundation}) --- exhibit similar mean MAE across all compositions $\left<\delta_{\t{MAE}}\right>$, as shown in fig.~\rref{fig:figure_2}{b} (see Supplementary Section 1 for more details).
These large errors reveal that none of the tested uMLPs can accurately resolve the fine energy differences associated with chemical disorder, highlighting a key shortcoming of compositionally agnostic MLP training. 
This motivates the development of strategies for designing training dataset that systematically and selectively sample the full range of chemical complexity present in solid solutions.

\subsection{Motif-based sampling for solid solutions}

\begin{figure*}[!htb]
  \centering
\includegraphics{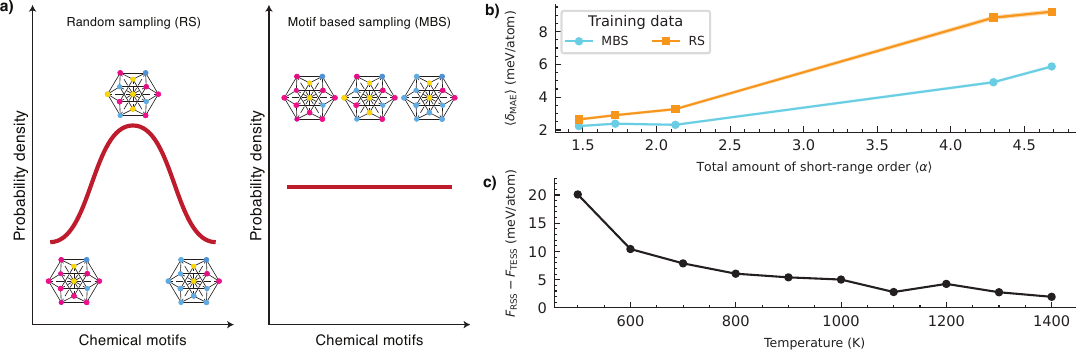}
\caption{\label{fig:figure_3} \textbf{Impact of motif based sampling (MBS) on the predictive performance of MLPs in solid solutions.} \textbf{a)} Illustration of the chemical motif population density in a random solid solution compared to that in an MBS dataset. \textbf{b)} Mean absolute energy error as a function of the total amount of short-range order (eq.~1) in the test set (averaged across the compositional space), highlighting that MBS-trained models consistently outperform random sampling. \textbf{c)} Free-energy difference between a CrCoNi random solid solution and its thermally equilibrated counterpart. The energy scale demonstrates that the accuracy gains provided by MBS are physically significant: they are necessary to resolve energetically competing configurations in alloy thermodynamics.  
  }
\end{figure*}

The accuracy of MLPs in modeling solid solutions hinges on how well their training datasets capture chemically relevant configurations --- a complicated task due to the overwhelming chemical complexity of solid solutions. Recently\autocite{sheriff2023quantifying, sheriff2024chemicalmotif, islam2024nonequilibriumchemicalshortrangeorder}, we demonstrated that this complexity can be systematically characterized by decomposing the structure into local coordination polyhedra --- called \textit{local chemical motifs} --- and analyzing their frequency and spatial correlations. Building on this framework, we introduce a \textit{motif-based sampling} (MBS) method to construct MLP training datasets that more uniformly represent local chemical environments, as illustrated in fig.~\rref{fig:figure_3}{a} (see the Methods section ``Sampling of chemical motifs'').

The impact of MBS on model performance is evaluated by comparing it against a baseline \textit{random sampling} (RS) approach, in which chemically random configurations are drawn uniformly across the ternary phase space (see the Methods section ``Training and testing datasets''). Both datasets contain identical numbers of structures, chemical compositions, and thermal noise levels. While the RS dataset reflects unaltered chemical randomness, the MBS approach refines these configurations via intracell atomic swaps that promote a more uniform distribution of chemical motifs --- reducing bias in motif frequencies and capturing a broader diversity of atomic environments.

Although both datasets preserve the overall chemical composition and maintain comparable degrees of chemical short-range order (SRO; see Supplementary Section 2), MBS achieves significantly higher \textit{motif packing density} --- defined as the percentage of unique motifs sampled in the dataset --- rising by 27\% in MBS relative to RS. Equivalently, the Jensen-Shannon divergence between the sampled and target uniform motif distribution is reduced by 0.0442\,bits in MBS (this divergence quantifies the similarity between two probability distributions, with lower values indicating a closer match to the target). This improvement scales with dataset size. For example, in a larger dataset with 702 configurations (compared to 66 configurations used to obtain fig.~\rref{fig:figure_3}{b}), the divergence drops from 0.4814 bits to 0.3661 bits, and the motif packing density increases by 38\%, such that 72.3\% of all possible motifs are sampled in the MBS dataset.

As a complementary approach, we also evaluated the widely used special quasi-random structure (SQS\autocite{SQS_1, SQS_2}) method, which is commonly employed to generate chemically disordered configurations for training machine-learned potentials. SQS constructs atomic arrangements that replicate the pair correlations of an ideal random alloy using optimization techniques. To assess its effectiveness, we applied SQS to each configuration in the 702-structure RS dataset. However, this did not lead to improved results: the Jensen–Shannon divergence slightly increased from 0.4814 bits to 0.4857 bits, and the fraction of unique chemical motifs remained approximately constant at 52.2\%.

To quantify how the improved motif diversity of MBS training sets affects model accuracy, we trained an ensemble of 10 MLPs per dataset (see the Methods section ``MLP model training'') and evaluated them on five test sets constructed with controlled degrees of SRO. These test sets were generated using reverse Monte Carlo (see the Methods section ``Training and testing datasets''), with the total amount of SRO quantified by the metric in eq.~1. Since all training datasets share identical thermal noise and lattice parameters, any differences in model performance can be attributed specifically to motif representation. As shown in fig.~\rref{fig:figure_3}{b}, MBS-trained models consistently outperform their RS counterparts, with the performance gap widening as the amount of SRO increases. This difference ranges from sub-meV/atom in near-random regimes to as much as 5\,meV/atom in high-SRO configurations (fig.~\rref{fig:figure_3}{b}, see Supplementary Section 3 for more details).

This improvement is physically meaningful: typical energy differences associated with SRO in alloys are on the order of 10\,meV/atom, thus, this accuracy is necessary to resolve energetically competing configurations in alloy thermodynamics (fig.~\rref{fig:figure_3}{d}, see the Methods section ``Free energy'', see Supplementary Section 4). By increasing motif diversity, MBS enables finer energetic resolution across chemically distinct configurations, supporting more accurate prediction of order--disorder energetics and related thermodynamic quantities such as free energies.

\subsection{Completing the training set: structural and thermal components}

To achieve robust predictive performance, the training set must incorporate not only chemical diversity, but also structural and thermal variations. While MBS accounts for chemical complexity, two additional components --- namely, phase sampling and thermal perturbations --- are needed to ensure coverage of distinct lattice types and vibrational states. These components are briefly summarized below, see the Methods sections ``Sampling of thermal effects'' and ``Training and testing datasets'' for a full description.

Phase sampling is achieved by including metastable phases that differ in structure and chemical order. In practice, this is performed by applying MBS to solid solutions of different lattices (e.g., fcc, bcc, and hcp), and by including a variety of stoichiometric compounds (i.e., ordered intermetallics). These additions expose the model to variations in local coordination and long-range order that are absent from chemically disordered phases.

Thermal perturbations are introduced via synthetic thermal noise that mimics vibrational effects, and by thermal expansion. In practice, configurations are isotropically expanded to match target thermal expansion factors, and further perturbed by random atomic displacements and strain fields. These augmentations are applied uniformly across all phases to ensure smooth interpolation over the vibrational configuration space.

\begin{figure}[!tb]
  \centering
  \includegraphics{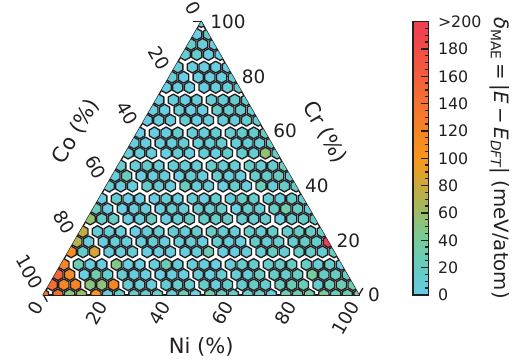}
  \caption{\label{fig:figure_4} \textbf{Energy prediction error across the CrCoNi composition space.} Mean absolute error (MAE) of MLP predictions using combined motif-based, phase, and thermal sampling. Compared to the universal MLP shown in fig.~\rref{fig:figure_2}{a}, this approach yields significantly lower and more uniform errors across the composition triangle.
  }
\end{figure}

We evaluate this comprehensive sampling strategy --- combining chemical, structural, and thermal components --- on the CrCoNi system. The resulting MLP achieves a substantial reduction in energy prediction error across the compositional space, as shown in fig.~\ref{fig:figure_4}. Compared to uMLPs (fig.~\rref{fig:figure_2}{a}), this approach produces significantly lower and more uniform errors across the full ternary composition triangle. The average MAE, shown in fig.~\rref{fig:figure_2}{b}, also confirms this improvement. \\

We now shift from developing and benchmarking the MBS framework to validating its predictive capabilities. To this end, we apply MBS to construct MLP training sets for additional systems, including the AuPt and AuCu binary alloys, as well as the high-entropy TiTaVW alloy. Validation is carried out across a range of experimentally accessible properties, including phase diagrams, chemical short-range order, thermoelastic and calorimetric behavior, and microstructural properties. Details of dataset construction, including DFT parameters and MBS parameters, are provided in the Methods sections ``Training and testing datasets'' and ``Density-functional theory''. For each system, we trained an MLP using the Performant Atomic Cluster Expansion (PACE\autocite{lysogorskiy_performant_2021}) model (see the Methods section ``MLP model training'').

\subsection{Phase diagrams}

\begin{figure}[!hbt]
  \centering
\includegraphics{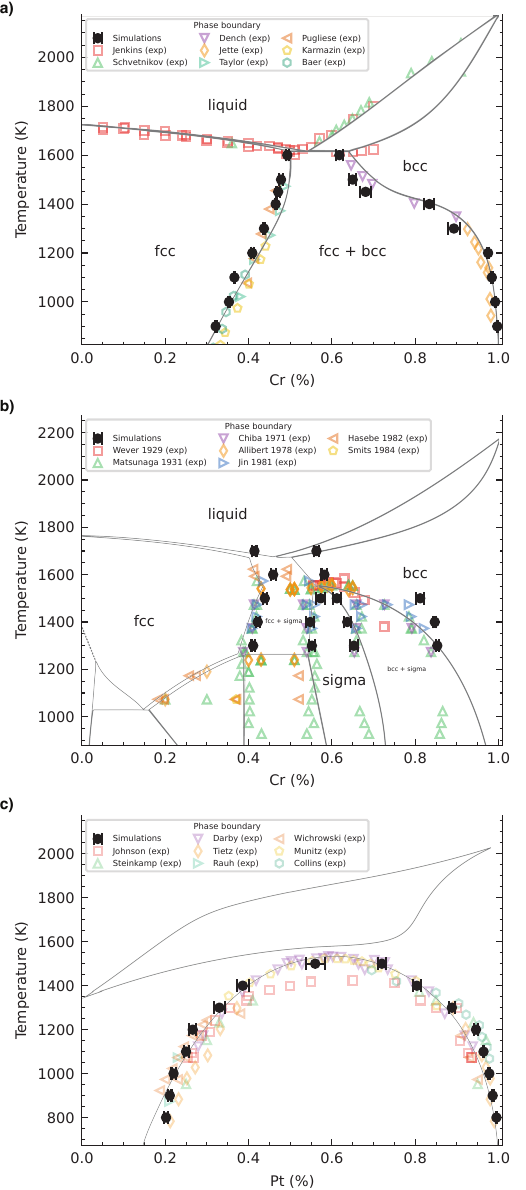}
\caption{\label{fig:figure_5} 
    \textbf{Phase diagram predictions for binary alloy systems.}
    Predicted solid-state phase boundaries along with corresponding experimental data\autocite{jenkins1937, svechnikov1962,dench1963, jette1934, taylor1951, pugliese1970, karmazin1978, baer1958, nash_crni_1986, wever1929, matsunaga1931, chiba1971, allibert1978, jin1981, hasebe1982, ishida1990, johansson1930, stenzel1931, darling1952, tiedema1957, raub1964, wictorin1949, munster1960, okamoto1985} and CALPHAD models\autocite{ASMhandbook, turchi2006, okamoto2003, okamoto1990} for the binary alloys: \textbf{a)} CrNi, \textbf{b)} CrCo, and \textbf{c)} AuPt. The CrNi and CrCo predictions were obtained using a single MLP trained on the ternary CrCoNi alloy. In contrast, the AuPt predictions were made using an independently trained binary MLP. 
  }
\end{figure}

To assess the thermodynamic fidelity of our MLPs, we evaluated their ability to reproduce experimental phase behavior in both binary alloys and compositionally complex high-entropy systems. This provides a stringent test of the model's accuracy beyond static energy prediction, requiring them to capture equilibrium thermodynamic transitions across a range of compositions, structures, and temperature conditions.

We first test our approach on the CrCoNi system. Using a single MLP trained on the full CrCoNi composition space (i.e., the model in fig.~\ref{fig:figure_4}) we computed phase diagrams for two of its binary subsystems: CrNi and CrCo. For all cases, solid-state phase boundaries were evaluated using the multi-cell Monte-Carlo approach\autocite{Niu2019}, modified to account for vibrational degrees of freedom (see the Methods section ``Phase diagrams''). The predicted phase diagrams are shown in figs.~\rref{fig:figure_5}{a} and \rref{fig:figure_5}{b}, where it can be seen that the MLP predictions closely match available experimental data and CALPHAD models, accurately capturing key phase boundaries. For example, in the CrNi binary the fcc-bcc boundaries align well with the experimental measurements, demonstrating the MLP’s ability to resolve competing crystal structures. To further illustrate the versatility of our approach, we also examined the AuPt system. The results are shown in fig.~\rref{fig:figure_5}{c}, where it can be seen that the model accurately captures the main feature of this phase diagram, namely the miscibility gap and its evolution with temperature. Notably, the miscibility gap marks a solid–solid transition where phase decomposition occurs within the same crystal lattice. This contrasts sharply with the fcc–bcc transition observed in CrNi (fig.~\rref{fig:figure_5}{a}), which arises from fundamentally different thermodynamic driving forces and exhibits distinct kinetic behavior. Across all systems, fig.~\ref{fig:figure_5} shows that the discrepancy between MLP predictions and experiment is comparable to the spread in experimental datasets themselves, indicating that the remaining differences are likely within the limits of experimental uncertainty.

\begin{figure}[!htb]
  \centering
\includegraphics{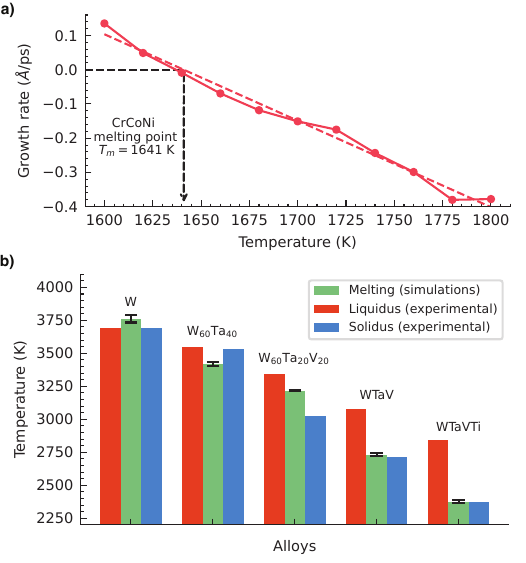}
\caption{\label{fig:figure_6} 
        \textbf{Predicted melting temperatures for high-entropy alloys.}
\textbf{a}) Solid–liquid coexistence simulation results for the equiatomic fcc CrCoNi alloy, with melting temperature identified at zero growth rate. The predicted melting temperature ($T_\text{m} = 1641\,\t{K}$) is within $3\%$ of the experimental value\autocite{exp_melting_pt}.
    \textbf{b}) Predicted melting temperatures for the bcc refractory high-entropy alloy TaTiVW and several derivative compositions, compared against experimentally reported solidus ($T_{\text{s}}$) and liquidus ($T_\ell$) temperatures\autocite{KIM2021113839}. Predictions for the alloys typically fall within the expected coexistence window between $T_{\text{s}}$ and $T_\ell$ (note that for pure elements $T_{\text{s}} = T_\ell$).
  }
\end{figure}

Having validated the model against well-characterized solid-solid transitions in binary alloys, we next evaluate melting points in chemically complex high-entropy alloys, for which phase diagram data is typically sparse or unavailable.
To evaluate model performance in this regime, we used the solid--liquid phase coexistence method\autocite{morris1994melting} to estimate melting points. This technique isolates a phase interface and tracks the steady-state growth rate of the solid phase as a function of temperature (see the Methods section ``Melting temperature''). For the equiatomic fcc CrCoNi alloy, the melting point was identified as the temperature at which the growth rate vanishes\autocite{karma} (fig.~\rref{fig:figure_6}{a}). The resulting estimate, $T_{\text{m}} = 1641$\,K, agrees within 3\% of experimental reports\autocite{exp_melting_pt} (note that the solidus--liquidus temperature gap in CrCoNi is small and has not yet been experimentally characterized). Because the liquid phase itself is chemically disordered, accurately predicting melting temperatures further demonstrates the MBS robustness in capturing chemical disorder beyond solid-state configurations.

We then applied the same procedure to the bcc TaTiVW refractory HEA and several of its derivative alloys (fig.~\rref{fig:figure_6}{b}). For each composition, the predicted melting temperature was compared with experimentally reported solidus ($T_{\mathrm{s}}$) and liquidus ($T_\ell$) values\autocite{KIM2021113839}. Our predictions are within 2\% for pure W, and they typically fall within the experimental $T_{\mathrm{s}}$--$T_\ell$ interval, as expected for the alloys\autocite{karma}: the melting point as measured in our simulations corresponds to the coexistence of solid and liquid phases, and should lie between the onset and completion of melting. 

Together, the results in fig.~\ref{fig:figure_5} and fig.~\ref{fig:figure_6} highlight the versatility and transferability of the MBS-trained MLPs. Without retraining or fine-tuning, the models generalize from chemically diverse training sets to accurately reproduce equilibrium phase behavior of binary subsystems and complex HEAs. This makes them a powerful tool for exploring phase stability and guiding design of new alloys across broad composition spaces.

\subsection{Chemical short-range order}

\begin{figure}[!t]
  \centering
\includegraphics{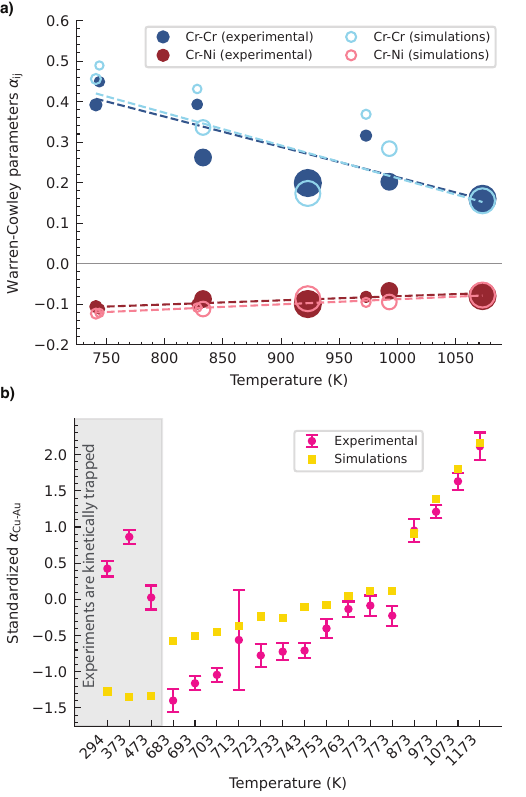}
\caption{\label{fig:figure_7} 
    \textbf{Short-range order predictions compared with experimental data.}
    \textbf{a}) Temperature-dependence of Warren–Cowley parameters for Cr$_x$Ni$_{(1-x)}$ alloys compared to diffuse-scattering experimental measurements\autocite{wc1,wc2,wc3} across the compositional space. The size of the data points are proportional to the Cr concentration, with $x$ ranging from 19.9\% to 33.3\%. Dashed lines are a linear fit to the data with corresponding color.
    \textbf{b}) Standardized Warren–Cowley parameters for the Cu$_{3}$Au alloy, comparing predicted values against experimental data derived from x-ray total scattering and reverse Monte Carlo modeling\autocite{OWEN201715}. Standardization highlights consistent qualitative trends despite  quantitative discrepancies attributed to known systematic energy-biases in the DFT functional\autocite{functionals_1,functionals_2,functionals_3}. Standard error of the mean of all simulation results are smaller than the data points.
}
\end{figure}

\begin{figure*}[!ht]
  \centering
\includegraphics[width=\textwidth]{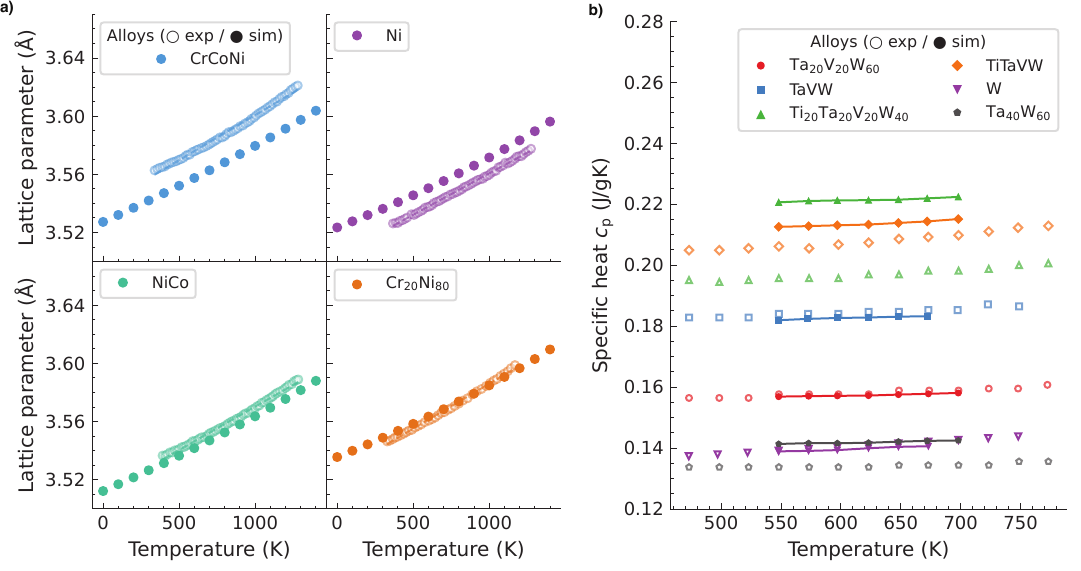}
\caption{\label{fig:figure_8} 
    \textbf{Thermophysical properties of CrCoNi and TaTiVW.} \textbf{a)} Temperature dependence of lattice parameters for CrCoNi and selected binary derivatives compared to experiments from ref.~\cite{JIN2017185}. \textbf{b)} Specific heat of TaTiVW-based alloys compared with experimental data from ref.~\cite{KIM2021113839}, except for pure W, which is compared to ref.~\cite{nist}. 
}
\end{figure*}

Chemical SRO is the tendency of solid solutions to deviate from perfect randomness, i.e., a measure of local chemical correlations. While solid solutions are chemically disordered phases lacking periodic chemical arrangements, they nonetheless exhibit subtle energetic biases toward certain local chemical configurations. These biases, captured quantitatively through Warren-Cowley parameters\autocite{cowley_approximate_1950}, fundamentally characterize the statistical fluctuations in local chemical environments and directly reflect underlying thermodynamic equilibria. However, typical training datasets generated from purely random configurations or SQS inherently miss these energetically significant correlations, limiting the accuracy of conventional MLPs\autocite{sheriff2023quantifying, cao2024capturing}.

To address this limitation, our MBS explicitly targets motif diversity, thus providing richer training datasets capable of capturing the delicate energetic gradients associated with SRO. To quantitatively assess the accuracy of our MLPs in predicting SRO, we computed the first-nearest-neighbor Warren–Cowley parameters, $\alpha_{ij}$ (eq.~2), from atomistic simulations of thermally equilibrated solid solutions (see the Methods section ``Warren-Cowley parameters''). 

Figure \rref{fig:figure_7}{a} compares the temperature dependence of Warren-Cowley parameters predicted for various Cr$_x$Ni$_{(1-x)}$ alloys against experimental values obtained from diffuse-scattering measurements\autocite{wc1,wc2,wc3} (with $x$ ranging from 19.9\% to 33.3\%). The close agreement demonstrates that our MBS training captures the nuanced energetic biases responsible for SRO across the compositional space. Similarly, fig.~\rref{fig:figure_7}{b} compares the predictions for the Cu$_3$Au alloy with experimental data from Owen et al.\autocite{OWEN201715}, who derived Warren-Cowley parameters from in situ x-ray total scattering and reverse Monte Carlo refinement. Although predicted magnitudes of $\alpha_{ij}$ differed slightly from experiments (see Supplementary Section 5) --- likely due to known systematic binding-energy biases of the DFT functional affecting lattice parameters and thus local ordering --- the experimental trends can be accurately captured by normalizing the Warren-Cowley parameters to correct for the strength of interactions, removing some of the DFT biases\autocite{functionals_1,functionals_2,functionals_3, Zeni2021}.

Overall, our results confirm that MBS effectively captures the subtle energetic biases that drive local chemical ordering, and that MLPs trained on such data can resolve SRO contributions. 

\subsection{Thermoelastic and calorimetric behavior}

To further evaluate the transferability of our MLPs, we investigated their capacity to capture temperature-dependent thermophysical properties---specifically, thermal expansion and constant-pressure specific heat $c_p$ across the CrCoNi and TaTiVW alloy systems.

Figure~\rref{fig:figure_8}{a} shows the thermal expansion of CrCoNi and selected binary derivatives (see the Methods section ``Thermal expansion''). The predicted lattice parameters align closely with experimental observations. For example, the Cr$_{20}$Ni$_{80}$ and NiCo systems exhibit a maximum deviations from experimental results of only 0.1\% and 0.3\%, respectively, across the entire temperature range. The coefficients of thermal expansion can be found in Supplementary Section 6.
This close agreement demonstrates the model's accuracy in capturing anharmonic vibrational effects and associated thermal expansion behavior in complex metallic alloy systems.

Figure \rref{fig:figure_8}{b} shows constant-pressure heat capacities for TaTiVW--based alloys (see the Methods section ``Specific heat''). 
The simulations reproduce the expected monotonic increase in $c_p$ with temperature, reflecting vibrational anharmonicity and mode excitation effects. Comparison with measurements from NIST\autocite{nist} and recent measurements on refractory high-entropy alloys\autocite{KIM2021113839} further validates the quantitative reliability of our predictions. The largest observed discrepancy from experimental data was 13\% for Ti$_{20}$Ta$_{20}$V$_{20}$W$_{40}$, while TiTaVW and Ta$_{40}$W$_{60}$ deviate by a maximum of 4\% and 6\% respectively, and the other three compositions have deviations of less than 2\%.

\subsection{Computationally-aided microstructure design}

The preceding sections have highlighted the accuracy and fidelity of MBS-trained MLPs in capturing diverse thermophysical and thermodynamic properties across alloy compositions. In this section, we illustrate how such computational accuracy could potentially support materials design and development, particularly in scenarios where experimental exploration is prohibitively challenging or expensive.

A notable example is the stacking-fault energy (SFE, $\gamma_\mathrm{SFE}$), which significantly influences the deformation mechanisms and thus the damage tolerance of fcc alloys. In particular, the CrCoNi alloy has attracted considerable attention due to its exceptional mechanical properties at low temperatures, attributed largely to its low SFE. This low SFE facilitates the activation of alternative deformation pathways such as mechanical twinning and transformation-induced plasticity\autocite{OH2024119349}, particularly at low temperatures when traditional dislocation slip mechanisms are insufficient\autocite{Liu2022}. However, despite its central role in deformation behavior, SFE has not traditionally served as a practical alloy design parameter. Experimental determination of SFE is notoriously challenging, requiring careful interpretation of fault structures via electron microscopy, or indirect inference methods. Furthermore, recent investigations have underscored the complexity and subtlety involved in reliably extracting accurate SFE values from experiments, highlighting potential pitfalls and uncertainties in current methodologies\autocite{Shih2021}.

\begin{figure}[!tb]
  \centering
\includegraphics{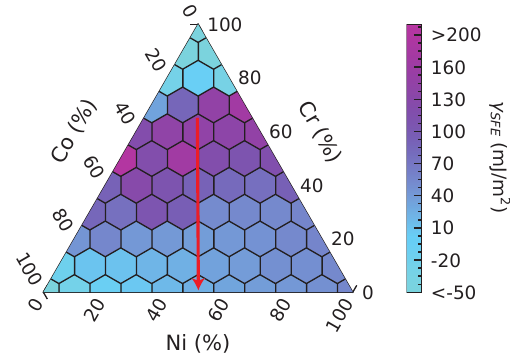}
\caption{\label{fig:figure_9} 
    \textbf{Computationally-aided microstructure design.} Predicted stacking-fault energy ($\gamma_{\t{SFE}}$) across the CrCoNi composition space at 500\,K. The arrow indicates the compositional path Cr$_x$Co$_{0.5-x/2}$Ni$_{0.5-x/2}$ (with $x \le 80\%$), along which SFE decreases significantly. Trends such as this one can inform materials design.
}
\end{figure}

To address this gap, we employed our MLP framework to systematically predict SFE across the entire compositional space of the CrCoNi alloy at 500\,K, explicitly accounting for thermal effects such as SRO at the appropriate length scales\autocite{cao2024capturing} (fig.~\ref{fig:figure_9}, see the Methods section ``Stacking-fault energy''). This computational approach captures subtle energetic contributions that are essential for accurate SFE prediction\autocite{cao2024capturing, Ding2018}. As shown in fig.~\ref{fig:figure_9}, the SFE exhibits a pronounced dependence on composition. Despite the complexity of this variation, clear trends emerge that can inform materials design. For example, reducing the Cr concentration along the Cr$_{x}$Co$_{0.5-x/2}$Ni$_{0.5-x/2}$ path (with $x \le 80\%$) leads to a marked decrease in SFE, as indicated by the arrow in \ref{fig:figure_9}. The influence of SRO on SFE across the compositional space is detailed in Supplementary Section 7, along with additional results at 800\,K (showing similar compositional trends).

These computational predictions combined with the predicted phase stability of fig.~\ref{fig:figure_5} illustrate how systematically exploring compositional dependencies of SFE can provide valuable insights for alloy design. By leveraging this predictive approach, one is able to efficiently identify promising alloy compositions, complementing and extending experimental capabilities in materials development.


\section{Discussion}

This work demonstrates that chemical motif diversity in the training dataset is a critical determinant of the predictive accuracy of MLPs for chemically disordered phases. The MBS strategy introduced here systematically improves model performance across compositions and structures by promoting a more uniform and physically meaningful representation of local chemical motifs. The success of this strategy hinges on recognizing that the fidelity of an MLP is limited not only by its architectural complexity or dataset size, but by the representational diversity of the atomic environments on which it is trained.

The generality of the MBS approach is evidenced by its consistent performance across a chemically diverse set of metallic alloys. These include systems with different magnetic, electronic, and structural complexity, all trained from datasets generated with the same consistent first-principles workflow. Importantly, MBS can be incorporated with negligible computational cost into existing training pipelines to enhance the performance of both MLPs and uMLPs without requiring model architectural changes. While random sampling or SQS has historically been the default in many MLP training protocols that account for disordered phases\autocite{li_complex_2020, zheng_multi-scale_2023, yin_atomistic_2021}, our results demonstrate that MBS offers a more robust and physically informed alternative, and should be preferred in applications where transferability and error control are critical.

Recent uMLPs architectures---particularly those trained on chemically diverse datasets---show promise for accurately capturing the properties of materials across the compositional space\autocite{barrosoluque2024open, shuang2025universal, ZHU2025120747, levine2025openmolecules2025omol25}. However, the sheer breadth of their training data often obscures the origins of their performance, underscoring the need for systematic validation. In this context, the MBS training and test datasets introduced here provide a valuable benchmark for evaluating the applicability of uMLPs across the compositional space.

The ability to reliably compute key materials properties across compositions --- from phase stability and thermophysical behavior to microstructure energetics --- hinges on the availability of accurate and efficient interatomic models. Within this context, MBS improves the practicality of MLPs for alloy design by enhancing accuracy at fixed training-set size, thereby reducing the data burden typically associated with high-throughput property predictions. These predictions are central to materials development, for example, accurate free-energy predictions allow the construction of phase diagrams that inform heat treatment protocols, while SFE guide assessments of deformation mechanisms. When combined with uncertainty quantification, such predictions provide actionable guidance for materials development. A promising future direction is the development of automated frameworks for high-throughput phase diagram generation. Although our current approach achieves accurate predictions of phase boundaries, it still relies on prior knowledge of the relevant competing phases. Generative models\autocite{Zeni2025} offer a potential solution by autonomously proposing both stable and metastable structures across compositional and structural landscapes. 

It is essential to acknowledge, however, that the ultimate accuracy of the predicted properties remains bounded by the quality of the reference data used to train the MLPs. The exchange-correlation functionals employed in DFT calculations in this work have well-known limitations in their predictive accuracy\autocite{functionals_2,functionals_3}. Nevertheless, functional development remains an active area of research, with recent advances (e.g., r2SCAN\autocite{r2scan} and ML density functionals\autocite{cider, bandgap,akashi2025machineslearndensityfunctionals}) offering improvements in physical accuracy. Higher-fidelity quantum chemistry methods may provide an even more reliable foundation for MLP training sets\autocite{chen}.

Crucially, even if a perfect electronic structure method was available --- capable of producing exact quantum mechanical results --- the methodological framework introduced here would remain indispensable. High-accuracy reference data alone does not guarantee generalization across chemical composition space, nor does it obviate the need for thoughtful sampling of chemical disorder during training. The MBS approach offers a systematic path toward generalizable and practically useful MLPs. Moreover, our findings suggest that the current generation of functionals may already be sufficient for property predictions of \textit{practical} relevance, so long as the compositional and configurational diversity of the training data is carefully managed.

In conclusion, this work advances the development of MLPs by addressing two persistent limitations: the absence of systematic strategies for chemical sampling of disordered phases and the difficulty of maintaining accuracy across broad compositional ranges. By introducing a chemical-motif-based framework for training set construction and demonstrating its effectiveness across multiple alloy systems, we provide a foundation for building MLPs that are both accurate and robust --- delivering reliable predictions across structures and compositions. This predictive performance, further supported by close agreement with experimental data across a range of thermodynamic and thermophysical properties, underscores the potential of MLPs to transition from niche research tools to a core component of multiscale simulation pipelines in materials design.

\printbibliography[heading=bibnumbered,title={References}]

\end{document}